%% file: main.tex
\documentclass[sigconf]{acmart}
\usepackage{tabularx}
\usepackage{graphicx}
\usepackage{multirow, multicol, makecell}
\usepackage{enumitem}
\usepackage{url}
\usepackage[normalem]{ulem}  % for \sout
\usepackage{soul}

\copyrightyear{2026}
\acmYear{2026}
\setcopyright{cc}
\setcctype{by}
\acmConference[WSESE '26]{3rd International Workshop on Methodological Issues with Empirical Studies in Software Engineering }{April 12--18, 2026}{Rio de Janeiro, Brazil}
\acmBooktitle{3rd International Workshop on Methodological Issues with Empirical Studies in Software Engineering (WSESE '26), April 12--18, 2026, Rio de Janeiro, Brazil}
\acmPrice{}
\acmDOI{10.1145/3786149.3788306}
\acmISBN{979-8-4007-2382-7/2026/04}


\title{OLAF: Towards Robust LLM-Based Annotation Framework in Empirical Software Engineering}

\author{Mia Mohammad Imran}
\affiliation{%
  \institution{Missouri University of Science and Technology}
  \city{Rolla, Missouri}
  \country{USA}
}
\email{imranm@mst.edu}

\author{Tarannum Shaila Zaman}
\affiliation{%
  \institution{University of Maryland Baltimore County}
  \city{Baltimore, Maryland}
  \country{USA}
}
\email{zamant@umbc.edu}

\begin{document}

\begin{abstract}
Large Language Models (LLMs) are increasingly used in empirical software engineering (ESE) to automate or assist annotation tasks such as labeling commits, issues, and qualitative artifacts. Yet the reliability and reproducibility of such annotations remain underexplored. Existing studies often lack standardized measures for reliability, calibration, and drift, and frequently omit essential configuration details. We argue that LLM-based annotation should be treated as a measurement process rather than a purely automated activity. In this position paper, we outline the \textbf{Operationalization for LLM-based Annotation Framework (OLAF)}, a conceptual framework that organizes key constructs: \textit{reliability, calibration, drift, consensus, aggregation}, and \textit{transparency}. The paper aims to motivate methodological discussion and future empirical work toward more transparent and reproducible LLM-based annotation in software engineering research.
\end{abstract}


\maketitle

%% ================================

\input{intro}

\input{problems}

\input{olaf/olaf}
\input{guideline}

\input{conclusion}
% =============================================================

\balance
\bibliographystyle{ACM-Reference-Format} 
\bibliography{llm-annotation}
\end{document}

%% file: intro.tex
\section{Introduction}

Empirical software engineering depends on labeled data, such as issues identified as bugs, code comments marked as self-admitted technical debt, or developer commit messages classified by intent \cite{Guéhéneuc2019}.  
Traditionally, producing such annotations has required substantial human effort. Manual annotation, though considered as gold standard, is labor-intensive and can be inconsistent~\cite{imtiaz2018sentiment}.
The rise of large language models (LLMs) has transformed this process. In recent years,
Researchers have employed models such as GPT, Claude, or Llama to generate preliminary labels or even to replace manual annotation entirely~\cite{gilardi2023chatgpt, churchill2025gpt, ahmed2025can}.

While this approach accelerates data production, it also introduces new methodological concerns.  
When an LLM assigns a label, it effectively functions as a \textit{measurement instrument}.  
Just as code metrics must be validated before use, labels produced by LLMs need to be tested for reliability and reproducibility.  
As noted by Wagner et al., many SE studies involving LLMs neglect to report standard measures of reliability or calibration~\cite{wagner2025towards}.  
Baumann et al. observe that minor configuration choices, such as prompt phrasing or sampling temperature, can completely reverse outcomes,  a form of prompt drift -- which they refer as \textit{LLM hacking}~\cite{baumann2025hacking}.  
Ahmed et al. find that LLMs can perform on par with human annotators, but only under carefully controlled evaluation conditions~\cite{ahmed2025can}. Meanwhile, Wang et al. empirically studied LLM-as-a-Judge in SE tasks and noted that LLMs could potentially substitute human evaluations in certain SE tasks while they may be provide  satisfactory and consistent performance in other tasks~\cite{wang2025can}.

These findings highlight a key gap: the lack of operationalization—the formal link between abstract concepts (e.g., “label accuracy”) and measurable procedures~\cite{haucke2021numbers}. Without clear operational definitions for constructs like reliability, consensus, or drift, ESE cannot guarantee reproducibility. While existing guidelines, such as those by ACM and Baltes et al.~\cite{baltes2025guidelinesempiricalstudiessoftware}, offer general advice on adopting LLMs in SE, they do not specifically address the need for operationalizing annotations in SE tasks. To fill this gap, we introduce the \textit{Operationalization for LLM-based Annotation Framework (OLAF)}, a structured approach that treats LLM-based annotation as a measurable and auditable process. To lay the groundwork for OLAF, we first explore why operationalization is essential and how current practices fall short in ensuring methodological rigor. 

\smallskip
\noindent
\color{blue}
\emph{
We outline detailed guideline and examples of OLAF here: \color{red}\url{https://se-llm-annotation-olaf.github.io/olaf/}
}
\color{black}

%% file: problems.tex
\section{Need for Operationalization in LLM-based Annotation}

% SE tasks increasingly rely on LLM-generated annotations, but the process of defining, validating, and reporting them remains underdeveloped. 

\subsection{Fragmented Annotation Practices}
The challenge of using LLMs for annotation lies not in performance but in methodological rigor. Many studies lack clear label definitions and consistent operationalization of reliability, calibration, or consensus, and often omit key configuration details (e.g., temperature). Based on recent evaluations~\cite{wagner2025towards, baumann2025hacking, ahmed2025can, wang2025can}, we summarizes these weaknesses and their implications.

% Although LLMs are widely applied in SE research, there is limited consensus on how to report or validate model-assisted annotation. 
\subsubsection{Missing Details}
Ahmed et al.~\cite{ahmed2025can} noted that even when models reach human-level accuracy, details such as prompt templates, model versions, and decoding parameters are sometimes omitted. 
Without these details, replication becomes uncertain. 
Several patterns appear from recent observations~\cite{ahmed2025can,baumann2025hacking,wagner2025towards}:

\begin{itemize}[leftmargin=*]
  \item \textbf{Limited reliability reporting.} Some studies provide only raw accuracy without chance-corrected measures (e.g., Cohen's~$\kappa$).
  \item \textbf{Incomplete configuration description.} Prompt templates, model versions, and decoding parameters are often underspecified.
  \item \textbf{Minimal calibration assessment.} Model confidence and stability across versions are rarely examined. For example, researchers reported the performance difference between different GPT model versions during annotation~\cite{aldeen2023chatgpt}.
\end{itemize}

\subsubsection{Configuration Sensitivity}
Baumann et al.~\cite{baumann2025hacking} found that minor configuration changes, such as prompt wording, temperature, or model version, can alter results in text-annotation tasks. 
This sensitivity complicates the expectation of stable measurement conditions. 
In practice, it appears as:
\begin{itemize}[leftmargin=*]
  \item \textbf{Prompt variation:} minor lexical variations (e.g., ``bug fix'' vs. ``defect repair'') yield different outputs. The problem of \textit{Prompt Perturbation}~\cite{chatterjee-etal-2024-posix} is still an under-explored area in general;
  \item \textbf{Model variation:} API updates modify model behavior under the same identifier;
  \item \textbf{Sampling variation:} stochastic decoding introduces variability across runs.
\end{itemize}

\subsubsection{Unstandardized Constructs and Metrics}
ESE commonly employs metrics for inter-rater reliability and measurement validity. 
In contrast, LLM-based studies rarely define comparable constructs for annotation quality. 
Ahmed et al.~\cite{ahmed2025can} propose human-model agreement ($\kappa_{H\text{-}M}$) and model-model consensus ($\rho_{M\text{-}M}$) as candidate indicators, while Baumann et al. discuss model sensitivity to configuration~\cite{baumann2025hacking}. 
% Despite those exploratory studies, no standard definitions have yet emerged.

% \subsection{Aggregation Across Multiple LLMs}
% When multiple LLMs or configurations assign labels to the same artifacts, their outputs may diverge. 
% To address this problem, a popular approach has been applying majority voting~\cite{tian2015max, badshah2025reference}, which treats each LLM as an annotator. Majority voting is particularly effective when LLM predictions exhibit high agreement or when limited overlap across annotators reduces the stability of more complex inference methods.
% A complementary class of techniques relies on probabilistic aggregation, including Dawid–Skene~\cite{dawid1979maximum}, GLAD~\cite{whitehill2009whose}, and MACE~\cite{hovy2013learning}, which infer latent consensus labels while estimating annotator reliability. These methods can capture systematic biases and asymmetric error patterns across LLMs but require sufficient annotator overlap and iterative inference, making them more resource-intensive and less interpretable. Although widely adopted in NLP~\cite{paun2018comparing}, probabilistic methods remain underused in ESE despite offering a principled way to handle disagreement among LLM annotators.

\subsubsection{Consequences for Research Validity}
The absence of operationalization undermines the validity of results across all levels of empirical reasoning. Construct validity suffers when labels lack defined meaning; internal validity is threatened by uncontrolled configuration sensitivity; and external validity is compromised by opaque or evolving model configurations that hinder replication.

\subsection{Problem Summary}
LLM-based annotation has advanced rapidly, but its methodological foundations remain underdeveloped. Existing studies follow fragmented practices, report reliability inconsistently, and rely on loosely defined constructs, making results difficult to interpret or reproduce. As a result, LLM-generated labels are often treated as measurements without being properly validated. Recent methodological analyses~\cite{ahmed2025can, baumann2025hacking, wagner2025towards} highlight the need to reconceptualize annotation as a structured measurement process rather than a purely automated task. Moreover, when annotation targets lack a clear gold standard or involve irreducible expert disagreement, reliability and calibration become ill-defined, rendering probabilistic outputs unsuitable for safety-critical or compliance-sensitive settings. Addressing these challenges requires standardized constructs, transparent configuration reporting, and reproducible aggregation methods. 
In the following section, we introduce \textit{OLAF}, a framework designed to meet these requirements.

%% file: olaf/olaf.tex
\section{Operationalization for LLM-Based Annotation
Framework (OLAF)} 

We provide detailed description of the proposed framework below.

\input{olaf/measurement}

\input{olaf/context}

%% file: olaf/measurement.tex
\subsection{Measurement of LLM Annotation}\label{sec:measurement}

The methodological issues that we discussed earlier, fragmented practices, configuration sensitivity, and undefined constructs, \textul{stem from treating LLM annotation as an \textit{automation task} rather than as \textit{measurement}}. 
When a model assigns a label, it acts as a measurement instrument whose reliability must be empirically established.
% , similar to how traditionally metrics are measured in annotation SE tasks.

\input{tables/table_metrics}

Recognizing annotation as measurement provides a foundation for explicit operational constructs and reproducible procedures. 
Following Ahmed et~al.~\cite{ahmed2025can} and Wagner et~al.~\cite{wagner2025towards}, 
we define six key dimensions that characterize LLM annotation as a measurable process: 
\textit{reliability, consensus, calibration, drift, aggregation}, and \textit{transparency}. 
Table~\ref{tab:olaf} summarizes them. Each specifies a property of measurement that can be operationalized and reported. We note that
\textit{these constructs presuppose that the target construct is observable and supports a stable measurement interpretation.
When this prerequisite is not met, the proposed operationalization may not be suitable for the task.} 
Below we describe the six metrics that form the conceptual and operational basis of \textit{OLAF}:

\begin{enumerate} [leftmargin=*]

\item \textbf{Annotator Reliability}~\cite{cohen1960} measures agreement across individual annotators, human or model. Chance corrected statistics such as Cohen’s $\kappa$~\cite{cohen1960}, Gwet’s AC1~\cite{gwet2014handbook}, and Krippendorff’s $\alpha$~\cite{krippendorff2018} quantify consistency beyond random alignment.

\item \textbf{Consensus}~\cite{buyl2025ai} reflects group level agreement, indicating whether annotators converge on a shared judgment or expose ambiguity in the task. Correlation based metrics capture the strength of this collective alignment.

\item \textbf{Aggregation}~\cite{li2025dna} combines multiple annotators into a single label. We outline two approaches: \textul{i) probabilistic aggregation}, which infers latent labels by modeling annotator error patterns (e.g., Dawid–Skene, GLAD, MACE)~\cite{he2024if}; and \textul{ii) majority voting}, suitable when agreement is high and multiple LLMs have comparable reliability~\cite{snow2008cheap}.

\item \textbf{Transparency}~\cite{liao2023ai} concerns documentation quality, ensuring reproducibility through complete disclosure of model identifiers, parameters, prompts, and settings, following practices such as Datasheets and Model Cards~\cite{oreamuno2024state}.

\item \textbf{Calibration}~\cite{geng2024survey} evaluates probability correctness by comparing predicted confidences with empirical accuracy. Metrics such as expected calibration error and Brier Score quantify miscalibration and indicate how trustworthy model probabilities are when weighting or aggregating labels~\cite{wright2025aggregating}.

\item \textbf{Drift}~\cite{abdelnabi2025get} captures stability, measuring how much model behavior changes under prompt or context variation. Drift is quantified using activation distances (L2 or cosine), output distribution metrics such as Jensen-Shannon divergence (JSD), or prompt sensitivity indices like POSIX~\cite{chatterjee-etal-2024-posix}.

\end{enumerate}

% Together, these six constructs form the conceptual and operational basis of \textit{OLAF}, providing the vocabulary and methodological scaffolding needed to evaluate annotation as a measurable and auditable process.

\textit{\textul{Note that not all six dimensions may not be required for every Software Engineering annotation task.} 
% Their relevance depends on the nature of the construct being measured, the stability of model outputs, and whether multiple annotators or model configurations are involved.
} For example, some SE tasks that rely on subjective judgments (e.g., \textit{sentiment/emotion analysis}, \textit{bug severity identification}) may benefit from reporting reliability or consensus, while tasks with more deterministic outputs (e.g., \textit{API extraction}, \textit{test-input generation}) may not require all dimensions. Practitioners should select the dimensions that meaningfully characterize the measurement properties of the specific task rather than applying the full set by default.

%% file: tables/table_metrics.tex
\begin{table*}[t]
\caption{Operationalization for LLM-based Annotation Framework (OLAF)}
\label{tab:olaf}
\footnotesize
\begin{tabular}{|l|p{3cm}|p{5cm}|l|}
\hline
\textbf{Construct} & \textbf{Measures} & \textbf{Metric / Indicator} & \textbf{Reporting Focus} \\
\hline
Annotator's Reliability~\cite{cohen1960}   
  & Annotator agreement 
  & Cohen's $\kappa$~\cite{cohen1960}, 
  Gwet’s AC1~\cite{gwet2014handbook}, Krippendorff’s $\alpha$~\cite{krippendorff2018} 
  & Stability of individual annotators \\
\hline 

Consensus~\cite{buyl2025ai}     
  & Group agreement 
  & Correlation metrics 
  & Collective alignment \\
\hline 

Aggregation~\cite{li2025dna}   
  & Combining annotators 
  & i) probabilistic~\cite{he2024if}: Dawid-Skene, GLAD, MACE; ii) Majority voting~\cite{snow2008cheap} 
  & Quality of fused labels \\
\hline 

Transparency~\cite{liao2023ai}  
  & Documentation quality 
  & Model (versions, parameter size, etc) and prompt details 
  & Reproducible annotation setup \\
\hline 

Calibration~\cite{geng2024survey}   
  & Probability correctness 
  & ECE, Brier Score~\cite{wright2025aggregating} 
  & Reliability of confidence values \\
\hline 

Drift~\cite{abdelnabi2025get}         
  & Behavioral stability 
  & L2, cosine, JSD, POSIX~\cite{chatterjee-etal-2024-posix} 
  & Robustness to prompt variation \\
\hline

\end{tabular}
\end{table*}

%% file: olaf/context.tex
\input{tables/table_configuration}

\subsection{Context of Annotation Workflows}
To operationalize these constructs in practice, it is necessary to account for how LLMs are embedded in the annotation workflow. 
Below we discuss six common annotation configurations~\cite{ahmed2025can, baumann2025hacking, wagner2025towards}.
% The point of integration, whether the model assists human annotators, verifies their work, or replaces them entirely, shapes which aspects of reliability, calibration, or drift require measurement and how they should be reported. In this sense, operationalization is inseparable from workflow design: methodological rigor depends on understanding when the model intervenes, how its outputs are validated, and what level of human oversight remains. The following taxonomy outlines typical configurations observed in recent empirical software engineering studies~\cite{ahmed2025can, baumann2025hacking, wagner2025towards}.

% \medskip
\noindent
\textbf{Human-in-the-Loop (HitL)} configurations pre-label artifacts using a model and require human validation of uncertain or low-confidence outputs~\cite{monarch2021human}. This reduces manual effort while maintaining quality control, though over-reliance on model confidence can propagate unverified errors. Typical indicators include $\kappa$, ECE, and selective-review accuracy.

\noindent
\textbf{LLM/Model-in-the-Loop (MitL)} integrates one or more machine learning/large language models as intermediate agents in the annotation pipeline~\cite{bartolo2022models}. Models generate candidate labels, rationales, or validation feedback, which are reviewed by humans or auxiliary models. Key metrics include human-model agreement ($\kappa_{H\text{-}M}$), cross-model agreement ($\kappa_{M\text{-}M}$), prompt sensitivity index (POSIX), and intra-annotator variance.

\noindent
\textbf{Verifier-in-the-Loop (VitL)} is a configuration where a dedicated verifier model assesses the validity of LLM-generated annotations prior to human review~\cite{wang2024humanllm}. The verifier predicts a confidence score indicating whether each label should be accepted, based on inputs such as the source text and model-generated explanation. Typical indicators include verifier accuracy, ECE, and change in reliability ($\Delta\kappa$) after selective re-annotation.

\noindent
\textbf{Model/LLM-as-Initial-Filter} configurations employ a model to discard clearly irrelevant items before manual or HitL processing~\cite{choi2024multi}. This improves scalability and efficiency but risks coverage bias by omitting rare or ambiguous cases. Metrics include recall of relevant items, false-negative rate, and $\Delta\kappa$ across filter runs.

\noindent
\textbf{LLM-as-a-Judge (LLMaaJ)} configurations treat models as automated evaluators applying rubrics to generated outputs such as explanations or summaries~\cite{wagner2025towards, lubos2024leveraging, wang2025can}. These yield consistent rubric enforcement but are vulnerable to evaluation bias and rubric drift. Representative measures include human-LLM agreement, rubric adherence, and prompt-sensitivity index.

\noindent
\textbf{LLM-as-an-Annotator} configurations delegate full labeling responsibility to the model~\cite{ahmed2025can}. This maximizes throughput and internal consistency but removes interpretability and introduces dependence on model parameters and version stability. Core metrics include inter-model consensus ($\rho$), ECE, and drift ($\Delta\kappa$).

These configurations are summarized in Table~\ref{tab:configurations}.

%% file: tables/table_configuration.tex
\begin{table*}[t]
\caption{Configurations of LLM Participation in Annotation Workflows}
\label{tab:configurations}
\footnotesize
\begin{tabular}{|p{2.5cm}|p{2cm}|p{6cm}|p{5.5cm}|}
\hline
\textbf{Configuration} & \textbf{Human Role} & \textbf{Primary Benefit} & \textbf{Main Risk} \\ 
\hline
Human-in-the-Loop~\cite{monarch2021human} & High &
Reduces cost by auditing low-confidence model outputs. &
Confidence miscalibration; selective review bias. \\ \hline
LLM/Model-in-the-Loop~\cite{bartolo2022models} & Moderate &
Combines model suggestion and cross-model validation to improve consistency and speed. &
Anchoring bias; prompt drift; compounded error propagation. \\ \hline

Verifier-in-the-Loop~\cite{wang2024humanllm} & Moderate &
Prioritizes human review via automated verification of LLM-generated labels. &
Verifier bias; dependence on small gold label sets; explanation quality sensitivity. \\ \hline

LLM-as-a-Pre-screen-Filter~\cite{choi2024multi} & High &
Scalable pre-screening of large corpora before human annotation. &
Coverage bias; missing rare but important samples. \\ \hline

LLM-as-a-Judge~\cite{wang2025can} & Low &
Objective rubric-based evaluation of generated artifacts. &
Evaluation bias; rubric drift; lack of interpretability. \\ \hline

LLM-as-an-Annotator \cite{ahmed2025can} & None &
Fully automated dataset labeling at minimal cost. &
Reproducibility loss from hidden drift and version changes. \\
\hline
\end{tabular}
\end{table*}

%% file: guideline.tex
\section{Guidelines for LLM-Based Annotation in Empirical Software Engineering}

We provide guideline for applying \textit{OLAF} below.

\noindent
{\bf Declare the Annotation Configuration.}
Researchers should explicitly state how LLMs participate in the annotation workflow (e.g.,  HitL, MitL, VitL, LLM-as-a-Filter, and LLMaaJ), as this determines which reliability and calibration measures are meaningful.

\noindent
\textbf{Document Model and Prompt Details.}
Report the LLM provider, model version, decoding parameters, and full prompt templates. Even minor prompt variations (e.g., `bug' vs.\ `defect') can affect annotation outcomes in SE tasks.

\noindent
\textbf{Aggregation and Reliability.}
When multiple LLMs are used, they should be treated as independent annotators and aggregated based on observed agreement. Majority voting or probabilistic aggregation can be applied. For example, in self-admitted technical debt detection, labels produced by different LLMs are compared to assess inter-model agreement. If agreement exceeds the predefined threshold, majority voting is used; otherwise, probabilistic methods such as Dawid-Skene are applied to infer latent labels while accounting for systematic differences between models.

\noindent
\textbf{Calibration and Drift Tracking.}
For SE tasks where LLMs output confidence scores, calibration should be evaluated. Drift should be monitored using a fixed, task-specific calibration subset. For example, consider issue classificaton annotation, a calibration subset of 50-100 issues annotated by at least two experts with substantial agreement (e.g., Cohen’s $\kappa \geq 0.6$) is retained. When the model version, prompt, or decoding parameters change, the LLM is re-applied to this subset and drift is quantified using changes in agreement ($\Delta\kappa$) and label distributions (e.g., Jensen-Shannon divergence). Another example, in a bug severity or priority annotation task, calibration is assessed by comparing model confidence scores against expert labels using Brier score or expected calibration error. Poor calibration indicates that confidence values should not be used for selective review or weighted aggregation.

%% file: conclusion.tex
\section{Limitations.}
OLAF treats LLMs as measurement instruments with quantifiable but limited stability. This assumption is necessarily approximate, particularly for proprietary models such as GPT and Claude, whose training data, optimization processes, and update schedules are opaque, and whose API versioning does not ensure parameter immutability. Consequently, calibration, drift, and reliability metrics can only bound observable variability.
Open-weight and open-source models, such as, Mistral, Llama, and Qwen, reduce some of these uncertainties through controlled inference settings and fixed checkpoints, but they still exhibit stochastic decoding, prompt sensitivity, and configuration-dependent variance. OLAF therefore adopts a notion of constrained stability, focusing on measurable regularities rather than deterministic or fully specified behavior.

\section{Conclusion and Future Plan}
% LLMs are increasingly used in empirical software engineering, yet their methodological grounding remains limited. This paper proposed \textit{OLAF}, a conceptual framework that defines six measurable constructs: reliability, calibration, drift, consensus, aggregation, and transparency, to treat LLM-based annotation as a reproducible measurement process.

% \textit{OLAF} is an initial proposal. Our future work will empirically evaluate its effectiveness against existing human and model annotation practices to determine whether it improves transparency, reproducibility, and comparability in LLM-based empirical software engineering research. We also intend to further improve \textit{OLAF} by formalizing boundary conditions for tasks where LLM-based annotation is unsuitable, refining the framework’s assumptions to better reflect LLMs as socio-technical systems subject to model and API variability, and developing SE-specific operational examples such as calibration subsets, drift-check procedures, and illustrative annotation workflows.

LLMs are increasingly used in empirical software engineering, but their methodological grounding remains limited. This paper introduced \textit{OLAF}, a framework that treats LLM-based annotation as a reproducible measurement process through six constructs: reliability, calibration, drift, consensus, aggregation, and transparency. \textit{OLAF} is an initial step. Our future work will empirically evaluate its impact on reproducibility and transparency, refine its assumptions for socio-technical LLMs, and develop SE-specific operational examples such as calibration subsets and drift-check procedures.